\documentclass[12pt]{iopart}
\usepackage{graphicx}
\usepackage{epstopdf, epsfig, rotating, floatpag, tikz}

\usepackage{subcaption}
\usepackage[noadjust]{cite} 
\usepackage{bm}
\usepackage{hyperref}
\usepackage{cleveref}
\usepackage{braket}
\usepackage{todonotes}
\usepackage{orcidlink}
\usepackage{amssymb}

\begin{document}

\title{The Widom line in the Ising model on a decorated bilayer lattice}

\author{Joseph Chapman$^1$\, \orcidlink{0009-0000-7737-5837}, Justas Gidziunas$^1$,  Bruno Tomasello$^{2,3,1}$\,\orcidlink{0000-0002-1156-5408}, Sam Carr$^1$ \,\orcidlink{0000-0001-9995-4944}}

\address{$^1$ Physics of Quantum Materials group, School of Engineering, Mathematics and Physics, University of Kent,  Canterbury, CT2 7NZ, UK }%
\address{$^2$ Dipartimento di Fisica e Astronomia `Ettore Majorana', Università di Catania, Via S. Sofia, 64, I-95123 Catania, Italy}
\address{$^3$ Functional Materials \& Sustainability Laboratory, Faculty of Forestry, Stefan Cel Mare University (USV), 13 University Rd, Suceava 720229, Romania}

\date{}

\begin{abstract}

There has been much recent interest devoted to a class of frustrated one-dimensional statistical mechanics lattice models which exhibit sharp thermodynamics. In this work, we study an extension of one of these models to two dimensions; the Ising model on a decorated bilayer lattice.
We show that the pseudo-transitions of the one-dimensional models become a real first order phase transition in this two-dimensional analogue. Moreover, the pseudo-transition behaviour is found to still exist above a bi-critical point. This can be characterised as a Widom line, which allows a re-interpretation of the physics in the previously studied one-dimensional models.

\end{abstract}

\noindent{\it Ising Model; Classical phase transitions; Frustrated systems; Spin chains, ladders and planes; Phase diagrams \/}

\maketitle

\section{\label{sec:intro}Introduction}
Phase transitions and critical phenomena have been the subject of constant study and fascination. Understanding critical phenomena often stems from study of various simplified mathematical models. Such an approach can provide precious insights about the essential and interesting physics without the complexities of real materials. The principle of universality \cite{wilson1983renormalization} ensures that the results from these simplified models can be directly applied to a variety of more realistic systems.

One such model, perhaps the most well known, is the Ising model \cite{1925Ising}, which provides a framework for the study of magnetic materials. In one dimension, as solved exactly by Ising over one hundred years ago, there is no phase transition at any finite temperature. This absence stems from the entropic benefit for disordering the system (through the creation of domain walls). In two dimensions, however, there is a phase transition at finite temperature, as shown by Onsager \cite{onsager1944crystal}. This transition is signalled by a discontinuity in one of the second derivatives of the free energy at some critical temperature, $T_c$. This non-analytic behaviour corresponds to the abrupt change as the system undergoes the transition. There is an associated order parameter, the spontaneous magnetisation, which is non-zero below $T_c$, and strictly zero above $T_c$. 
In contrast, the thermodynamics of (most) one-dimensional models are smooth, analytic functions of temperature, and there is no non-zero order parameter at any temperature $T > 0$. 

Recently, however, there has been a variety of one-dimensional models that exhibit curiously sharp thermodynamics at finite (non-zero) temperature \cite{chapman2024bifurcation,yin2023marginal,hutak2021low,strevcka2016spin,de2018quasi,carvalho2018quantum,rojas2019universality,rojas2019peculiarities,carvalho2019correlation,strecka2020pseudo,krokhmalskii2021towards,sznajd2022ising,rojas2025quantum,rojas2026dual,braz2025thermodynamic}. In these models, there is a rapid increase in the entropy at finite temperature, which is accompanied by a narrow peak in the specific heat. There are strict no-go theorems preventing any true phase transitions in one dimensional models with short-range interactions \cite{perron1907theorie,cuesta2004general}. Yet the aforementioned thermodynamics resemble those seen in higher dimensional models, albeit lacking a true singularity, which has motivated the search for an explanation. The one dimensional nature of these models typically permits an exact solution via the transfer matrix method, and thus allows a full mathematical understanding of the phenomena. 
This is highlighted in Ref.~\cite{rojas2020conjecture}, where Rojas provides a set of conditions on the ground state entropy as a function of some external parameter for which these "pseudo-transition" behaviours may be observed. Briefly, these conditions require that the entropy of two neighbouring phases be continuous from at least the one-sided limit. Satisfying this condition mathematically leaves traces of the pseudo-transition at finite temperature. Whilst these conditions are rigorous, what is missing is a deeper physical insight into the origin of these pseudo-transitions. 

The present work is motivated by the following two questions:
\begin{enumerate}
    \item What happens to these pseudo-transitions in higher-dimensional models, where \textit{real} phase transitions are possible? Do these pseudo-transitions give way to real phase transitions, or do they still exist somewhere in the phase diagram?
    \item Can we obtain deeper insights about the origin of these pseudo-transitions from a study of higher-dimensional models?
\end{enumerate}
To answer these questions, we study a natural generalisation of the classical Ising model on the ``Toblerone lattice" (a decorated two-leg ladder model, geometrically reminiscent of the iconic Swiss chocolate bar) \cite{chapman2024bifurcation} to higher dimensions in which the chains of Ising spins are replaced by square lattice planes of spins. We thus obtain a bilayer Ising model with both a direct and indirect coupling between the layers. This is illustrated in \Cref{fig:init_lattice_sketch}, and fully detailed in \Cref{sec_model}. This section provides further details of the model, and presents the equivalence to an undecorated bilayer, with a temperature dependent interaction between the layers. 

In \Cref{sec_critcal}, we derive the phase diagram, shown in \Cref{fig:phase_diagram}, as a function of temperature and one of the interaction parameters in the model. As one might expect, the high temperature phase paramagnetic, while at lower temperatures, there are two different ordered phases separated by a first order phase transition. This first order transition coincides with the location of the the pseudo-transition in the one-dimensional model. This first order line ends at a bi-critical point where it meets the phase boundary between the ordered and paramagnetic phases. However, we may extend the first order line into the paramagnetic phase, which is known as a Widom line. 

A Widom line was first discussed in the context of the phase diagram of water in Ref.~\cite{xu2005relation}. Here, the Widom line extends from the critical end-point of the liquid-gas transition and is (loosely) defined as the maximum (peak) of the specific heat capacity. In the supercritical fluid, the liquid and gas are the same phase, meaning no singular thermodynamics occur. However, close to the critical end-point, these peaks can become very narrow as they do indeed become singular across the first order phase transition on the other side \cite{mcmillan2010going,franzese2007widom,luo2014behavior}. The concept of the Widom line has since been extended beyond the liquid-gas transition, as studied by Widom, to the extension of \textit{any} first order line beyond its end-point. Examples include the Mott transition in the Hubbard model \cite{sordi2012pseudogap,fournier2025frenkel}, the quantum chromodynamics phase diagram \cite{sordi2024introducing}, and Fock-space properties of the extended Aubry-André-Harper chain \cite{roy2025many}.

In the context of this work, the Widom line manifests itself as a peak in the specific heat which becomes increasingly sharp as one approaches the bi-critical point. This necessitates a study of the thermodynamics of the model. In one-dimensional models, the thermodynamics can readily be obtained through the transfer matrix method, which permits painless analytical calculation of the partition function. Conversely, in two dimensions, exact analytical calculation is not usually possible. One exception is the single layer (square lattice) Ising model which is integrable \cite{onsager1944crystal}. In \Cref{sec_thermo}, we devise an approximation scheme around this single layer which allows us to study the thermodynamics (entropy and specific heat) of the decorated bilayer. There is a sharp, finite peak in the specific heat in these thermodynamics, corresponding to crossing the Widom line. This is reminiscent of the pseudo-transition seen in one-dimensional models. Drawing this parallel between the two allows us to offer a novel, alternative interpretation of the previously discussed pseudo-transitions as Widom lines, with the rest of the phase diagram suppressed due to the low dimensionality. This is discussed in \Cref{sec_widom}. We also find that this interpretation agrees with the existing conjecture of Ref.~\cite{rojas2020conjecture} for the existence of pseudo-transitions. This provides an additional, physical insight into the origins of these phenomena.

\section{\label{sec_model}The model}

\begin{figure}[tbp]
    \centering
    \includegraphics[width=\textwidth]{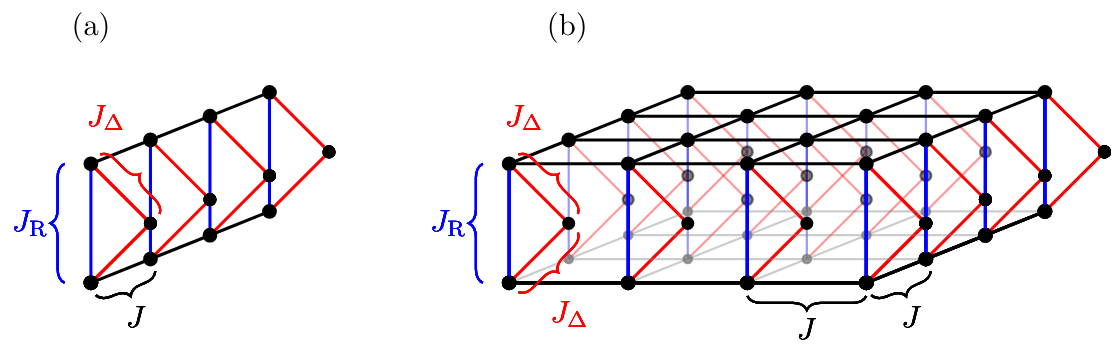}
    \caption{(a): The ``Toblerone lattice" studied in \cite{chapman2024bifurcation}. (b): The generalisation of the Toblerone lattice to two dimensions. This is the decorated bilayer lattice, defined by \Cref{model_hamiltonian}. Each circle is an Ising spin $s_i = \pm1$. The two square lattice layers are isotropic with an intra-layer interaction $J$ (black). These layers are coupled directly via $J_R$ (blue), and indirectly via $J_\Delta$ (red). We take $J_R < 0$ (AFM), and $J,J_\Delta >0$ (FM). Note that the decorating spins are not coupled to each other.}
    \label{fig:init_lattice_sketch}
\end{figure}

In the present work, we study a \textit{decorated} bilayer Ising model that provides a natural generalisation to two dimensions of the decorated one-dimensional model studied in Ref.~\cite{chapman2024bifurcation}. The model is defined by the Hamiltonian
\begin{equation}\label{model_hamiltonian}\hspace{-2em}
    \mathcal{H} = -\sum_{\langle i,j\rangle} J(s_i^{(1)} s_{j}^{(1)} + s_i^{(2)} s_{j}^{(2)}) - \sum_i \left(J_\Delta s_{i}^{(d)} (s_{i}^{(1)} + s_{i}^{(2)}) + J_\mathrm{R}(s_{i}^{(1)} s_{i}^{(2)}) \right), 
\end{equation}
where $\langle i,j\rangle$ indicates a sum over nearest neighbours, and the interactions are illustrated in \Cref{fig:init_lattice_sketch}. The spins $s_{i}^{(1)},s_{i}^{(2)}$ are those located in layer 1 and 2 respectively, and the index $i$ denotes the position within the layer. The spins that decorate each inter-layer bond are labelled as $s_{i}^{(d)}$. These planes are coupled by the same decorated bond that was considered in the one-dimensional model, that is to say that the spins in each plane are coupled to the corresponding spin on the opposite plane both directly (by $J_\mathrm{R}$), and indirectly (by $J_\Delta$). It is important to note that the spins $s_{i}^{(d)}$, that decorate each bond, are not coupled to each other.
Without loss of generality, the in-plane interaction $J$, and the coupling to the decorating spins $J_\Delta$ are taken to be positive (ferromagnetic)\footnote{For each of these interactions, there is a trivial duality transformation that changes the sign, but this is not needed for this work.}. More importantly, $J_R$ is taken to be negative (antiferromagnetic), which adds frustration to each triangle. 

\begin{figure}
     \centering
     \includegraphics[width=0.75\linewidth]{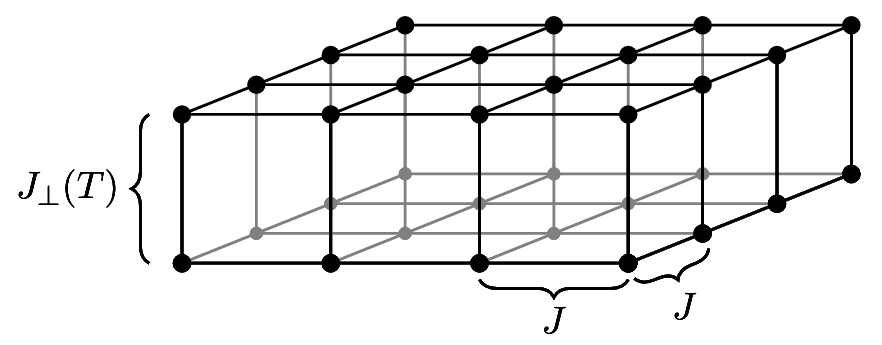}
     \caption{The effective bilayer model with the temperature dependent coupling between the layers that is obtained  following the exact summing out of the decorating spins in \Cref{fig:init_lattice_sketch}. This interaction $J_\perp(T)$ is defined in \Cref{RG_eq}.}
     \label{fig:rg_lattice}
 \end{figure}

Following the the same procedure as introduced for the one-dimensional case \cite{hutak2021low}, the decoration spins can be exactly traced out. We are able to relate the partition function $Z$, of the original decorated model, to the partition function $Z_b$ of the effective bilayer (without the decorating spins, illustrated in \Cref{fig:rg_lattice}): 
 \begin{equation}\label{rg_z}
     Z = \exp(C)\ Z_\mathrm{b}[J_\perp(T)].
 \end{equation}

This effective bilayer now has a temperature dependent coupling between the layers given by
 \begin{equation}\label{RG_eq}
 	J_\perp(T)=J_R + \frac{T}{2}\log(\cosh(2J_\Delta /T)),
 \end{equation}
and the additional parameter, $C$, is
 \begin{equation}\label{RG_const}
     C = \log(2) + \frac{1}{2}\log(\cosh(2J_\Delta /T)).
 \end{equation}

 Before we continue, it is worthwhile to summarise the behaviour of $J_\perp(T)$ as a function of temperature.  The crucial feature is that $J_\perp(T)$ changes sign as a function of temperature. This stems from the competition between the interactions $J_R$ and $J_\Delta$. For further details, the reader is encouraged to consult Refs.~\cite{hutak2021low,chapman2024bifurcation}.
 
The model in \Cref{model_hamiltonian} is thus reduced to a bilayer Ising model, albeit with a temperature dependent coupling between the layers. Whilst the bilayer Ising model is not integrable, and hence no exact expression exists for the partition function, many previous studies have set out to investigate the critical phenomena in these models \cite{ballentine1964critical,allan1970critical,abe1970some,lipowski1993layered,li2001criticalexponents,balcerzak2009magnetic,szalowski2012critical,szalowski2013influence,rossato2023ising,flynn2026moir,roos2024frustrated,diaz2017monte,moodie2020transition}. 
The introduction of an inter-layer coupling does not change the universality class from that of the standard 2D Ising one, but does give an important correction to the transition temperature. This is examined in more detail in the following section, along with the effects of having a temperature dependent $J_\perp(T)$.

\section{The Phase Diagram and Critical Behaviour}
\label{sec_critcal}
\subsection{Obtaining the phase diagram}
\begin{figure}
    \centering
    \includegraphics[width=0.8\linewidth]{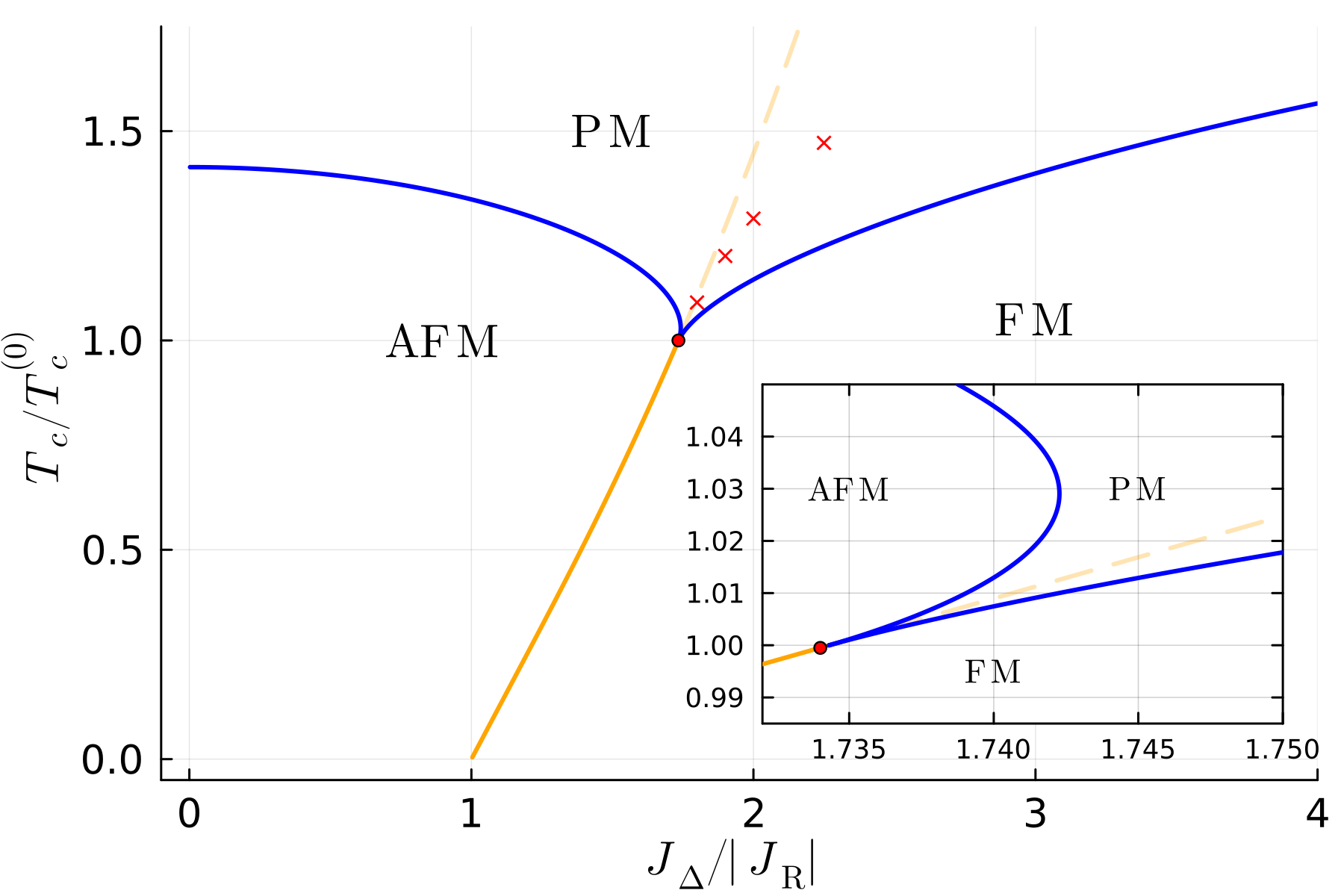}
    \caption{The phase diagram for the model in \Cref{model_hamiltonian}. The solid blue lines correspond to the lines of second order transitions, and the solid orange line corresponds to the line of first order transitions. This line separates two ordered phases; one in which the layers are oriented in the same direction (FM), and one in which they are oriented in opposite directions (AFM). Within each layer, the spins are ordered ferromagnetically ($J=1$). These lines meet at a bi-critical point, shown by the red circle. The ``$\times$" markers correspond to the observed peaks in the specific heat which is interpreted as a Widom line. The shift away from the line $J_\perp = 0$ (faint dashed orange line) is discussed in the main text. The inset shows a zoomed view near the bi-critical point, and highlights the existence of a re-entrant phase transition. The derivation of this phase diagram is discussed in the main text.}
    \label{fig:phase_diagram}
\end{figure}

In the previous section, we reduced the decorated model in \Cref{model_hamiltonian} to an effective bilayer model with a temperature dependent coupling between the layers $J_\perp(T)$. Let us now deduce the phase diagram of this model. We begin our discussion in the low temperature regime where the planes, in the absence of $J_\perp$, will each have a non-zero magnetisation. The task is then to consider the effect of non-zero $J_\perp(T)$. For $J_\perp >0$, the two layers tend to align their spins in the same direction, which we refer to as the \textit{ferromagnetic} phase (\textit{FM}). 
Similarly, for $J_\perp<0$, the layers will tend to align their spins in opposite directions, referred to as the \textit{antiferromagnetic} phase (\textit{AFM}). We stress here that this convention refers to the relative alignment of the spins between the two layers. The spins within the same layer are, at low temperature, ordered according to the intra-layer interaction $J >0$. Then it is quite clear that there must be a first order phase transition between these two phases as $J_\perp$ changes sign. This allows us to draw this line of first order transitions, corresponding to $J_\perp=0$, on the phase diagram in \Cref{fig:phase_diagram}.

We now turn to the discussion of the effect of the inter-layer interaction on the critical temperature between the ordered and disordered phases. The strong coupling limit $J_\perp \to \infty$ is considered first. In this case, it can be understood that the planes act in unison, and we obtain a critical temperature that is twice that of the individual layers $T_c = 2T_c^{(0)}$. The critical temperature of the single layer square lattice Ising model, $T_c^{(0)}$, in given in \ref{app_sq_latt}.

The other limit is for small $J_\perp(T)$. Trivially, for $J_\perp =0$, the critical temperature is just that of the Ising square lattice $T_c = T_c^{(0)}$. For non-zero values of $J_\perp$, the effect on the critical temperature is not as straightforward. A scaling relation is obtained through a perturbation expansion in $J_\perp$ in Ref.~\cite{abe1970some}, which is
\begin{equation}\label{tc_scaling_abe}
    T_c = T_c^{(0)}(1 + a\:| J_\perp|^{1/\gamma}),
\end{equation}
where the critical exponent $\gamma = 7/4$ for the 2D Ising model, and $a$ is a constant. We provide an alternative derivation of this relation in \ref{app_tc_scaling}. We emphasise that the exponent on $|J_\perp|$ is less than one. This means that as $J_\perp(T)$ goes through zero, the derivatives will be a discontinuous. Shortly, we will see that this feature manifests itself in the phase diagram, where the sub-linear dependence is dominant for small $J_\perp(T)$ (i.e. the region we are interested in). Whilst the effect on the critical temperature is singular, there is no change to the critical exponents of the bilayer. 

Now that we have established two limits for the critical temperature scaling as a function of $J_\perp$, we are able to write an empirical fitting function to interpolate between these limits. We write this function as
\begin{equation}\label{Tc_scaling_mc}
    T_c = T_c^{(0)}\left(1 + |\tanh(aJ_\perp)|^{4/7}\right),
\end{equation}
where $a$ is again a constant. In \ref{app_MC}, we demonstrate that this function is a good fit to Monte Carlo data, and full details of the Monte Carlo implementation and fitting can be found in \ref{app_MC}. From this fitting we are able to extract the value $a \approx 0.217 \pm 0.001$. Indeed, we deem this sufficient for the purposes of this present investigation. As a result of the temperature dependence of $J_\perp(T)$, we must solve \Cref{Tc_scaling_mc} self-consistently. From this, we are able to construct the lines of second order phase transition in \Cref{fig:phase_diagram}, which completes the phase diagram.

We have discussed the three phases, FM, AFM, PM, earlier in this section. 
The first order line, separating the FM-AFM phases, slopes towards the FM phase, indicating a lower entropy by the Clausius-Clapeyron relation \cite{carnot1872reflexions,reichl2016modern}. This can be understood by considering that in the AFM phase, the decorating spins become frustrated, yielding an entropy of $\log 2$ per spin, which is clearly larger than the constant entropy found in the FM phase.  

One may also observe that as temperature is \textit{increased}, and the first order line is crossed, the system remains ordered but in a different phase. This is understood through noting that the energetic penalty of the FM phase is offset by the entropic benefit afforded by allowing $s^{(d)}$ to point in either direction (become frustrated), thus we see order by disorder.  

The critical end point of this first order line meets the two lines of second order transitions at $T_c/T_c^{(0)}=1, J_\Delta / |J_{\mathrm{R}}| \approx 1.734$. Extending from this end point we identify a \textit{Widom line} \cite{xu2005relation}, as indicated by the (red) crosses in \Cref{fig:phase_diagram}. The full significance of this Widom line will be discussed in \Cref{sec_widom}, however we note that the existence of a Widom line in a lattice model is not unprecedented - see Refs~\cite{sordi2012pseudogap,fournier2025frenkel}, for example.

We now turn to the behaviour of the lines of second order transitions which has been touched on previously. The scaling behaviour of the critical temperature is dependent on the exponent on $|J_\perp|$ in \Cref{Tc_scaling_mc}. The sub-linear dependence has an important consequence. Namely, there is a range of values of $J_\Delta / |J_\mathrm{R}|$ for which there are multiple self-consistent solutions to \Cref{Tc_scaling_mc}. This can be seen through the curvature of the second order line in the inset of \Cref{fig:phase_diagram}. These multiple solutions lead to a re-entrant transition.

\subsection{Critical exponents}
\label{sec_crit_exp}
For the majority of the phase diagram, the critical exponents are unchanged from those of the 2D Ising universality class \cite{abe1970some}. At the bi-critical point, the critical exponents differ from those of the Ising square lattice. By using the critical temperature scaling in \Cref{tc_scaling_abe} with the scaling relations for the 2D Ising model, a new set of exponents can be obtained which are listed in \Cref{tab:Exponents}. A full derivation of these can be found in \ref{app_exponents}.
\begin{table}
    \centering
    \begin{tabular}{c|c|c}
         Exponent & Away from bi-critical point (2D Ising) & Bi-critical point  \\
         \hline 
         $\alpha$ & 0& 6/7\\
         $\beta$ & 1/8 & 1/14 \\
         $\gamma$ & 7/4 & 1 \\
         $\delta$ & 15 & 15 \\
        $\nu$ & 1 & 4/7 \\
        $\eta$ & 1/4 & 1/4 
    \end{tabular}
    \caption{Comparison of the critical exponents for the standard 2D Ising universality class (applicable to most of the phase diagram), and the exponents obtained at the bi-critical point. The coincidence of the exponents $\delta,\eta$ relates to the dependence of these exponents on a field. A discussion of the relevant universality class at the bi-critical point is given in \Cref{sec_crit_exp}.}
    \label{tab:Exponents}
\end{table}

From the pioneering work of Wilson on universality \cite{wilson1983renormalization}, we know that critical behaviour, such as the critical exponents, are determined by only the symmetry of the model and the number of dimensions. Other microscopic details of the model become irrelevant in the vicinity of this critical line. For the majority of the phase diagram, the only relevant symmetry is the $\mathbb{Z}_2$ of the Ising spins, which gives the Ising universality class. At the bi-critical point however, the coupling between the layers becomes zero exactly at the critical point. There is therefore an enhanced $\mathbb{Z}_2 \times \mathbb{Z}_2$ symmetry, as the two layers act independently at this point. In fact, according to the field theory, \Cref{eq:DSG} in \ref{app_exponents}, $g,\lambda$ both vanish at the bi-critical point, leaving the Gaussian model with an $O(2)$ symmetry. Whilst dynamical symmetry enlargement is not uncommon, it is not obvious how to interpret it in this case where the discrete $\mathbb{Z}_2 \times \mathbb{Z}_2$ symmetry becomes the continuous $O(2)$ symmetry. A discussion of this phenomenon and its consequences is left for future work. 

\section{Thermodynamics}
\label{sec_thermo}

In order to probe the thermodynamics of the non-integrable model in \Cref{model_hamiltonian}, we must make an approximation. We are not trying to describe the physics of the whole phase diagram here, but we aim to study the regions near the transitions (i.e. for small $J_\perp(T)$) which are of interest for the present study. To formulate this approximation, we begin from the relation for the partition functions in \Cref{rg_z}, and we obtain the free energy, $F$, in the usual way, which we write as 
\begin{equation}
    F = - T\log(2) + ( J_{\mathrm{R}} - J_\perp(T)) + F_b(J_\perp(T),T),
\end{equation}
where $F_b$ is the free energy of the bilayer at the value of $J_\perp$ at temperature $T$. Taking the derivative of this free energy gives the entropy as
\begin{equation}\label{entropy_relation}
    S = S_b + \left(1 - \frac{\partial F_b}{\partial J_\perp} \right)\frac{\partial J_\perp}{\partial T} + \log 2 \equiv S_b + S_\mathrm{dec.},
\end{equation}
where we can collect terms so that the entropy is separated into two parts. We have the entropy of the undecorated system, $S_b$ (at fixed $J_\perp(T)$), and the entropy that can be directly attributed to the decoration $S_{\mathrm{dec.}}$. This relation is exact, and the separation into these constituent parts provides the starting point for our approximation method. We calculate these two parts in two different ways.

First, we discuss the approach to $S_b$. Since the bilayer is not integrable, we cannot evaluate the partition function, nor can we calculate the free energy. As will become clear shortly, we only require that $S_b$ gives the logarithmic singularity (as is known occurs in the 2D Ising model) at the correct temperature, as defined by \Cref{Tc_scaling_mc}. Thus, we approximate this as 
\begin{equation}\label{S_bil_approx}
    S_\mathrm{b} \approx 2 S_0\left(\frac{T}{T_c} \; T_c^{(0)}\right),
\end{equation}
where $S_0$ is the entropy of the square lattice Ising model, as given by \Cref{sq_latt_entropy} in \ref{app_sq_latt}, and $T_c$ is defined through the semi-empirical relation in \Cref{Tc_scaling_mc}. Scaling the temperature in this way gives the logarithmic singularity in the correct place. Indeed, as shown in \Cref{tab:Exponents}, away from the bi-critical point, the bilayer exhibits 2D Ising universality, which is captured through the use of the exact result for the square lattice Ising model. The dominant fluctuations near to the second order transition, but away from the bi-critical point, are those within each plane which are well described by this approach to $S_b$. Close to the bi-critical point, however, this is clearly not sufficient, as evidenced by the differing critical exponents. As yet, we do not have an effective approximation directly through the bi-critical point, and so we do not plot thermodynamic quantities through here. 

We now turn to the other term $S_{\mathrm{dec.}}$, the calculation of which necessitates more discussion. This term will be responsible for the first order transition, and for the physics across the Widom line. Our approach to the calculation of this term is motivated by a result from the one-dimensional model in Ref.~\cite{chapman2024bifurcation}; the width of the peak in the specific heat was inversely proportional to the correlation length. The full details of this approximation are given in \ref{app_ICM}, but we give a brief summary here. We define a cluster of spins that has the size of the correlation length; we take the system size $N = \xi^2$, where we use the scaling form of the correlation length for the square lattice Ising model in \Cref{sq_latt_xi} in \ref{app_sq_latt}. Within this cluster, all of the spins act in unison. This means that we can reduce the whole cluster of spins to just a single, effective spin (in practice we restrict the partition function summation over only these states with $s_i^{(1)} = s_{j}^{(1)}, s_i^{(2)} = s_{j}^{(2)},  \ i,j \in \{ \xi \}$). This reduction allows evaluation of the partition function, and hence calculation of the thermodynamics. Henceforth we refer to this model as the ``Correlated Cluster Approximation", or \textit{CCA}. This approximation scheme is primarily motivated by the physics of the model, and provides some quantitative basis allowing the thermodynamics to be studied. Of course, these thermodynamics will not be exact but they will serve to illustrate the physics of the phase diagram in \Cref{fig:phase_diagram}. 

One essential feature of the model in \Cref{model_hamiltonian} is the effect of $J_\perp(T)$ changing sign as a function of temperature. In \Cref{sec_critcal}, we discussed that this change of sign can result in a first order phase transition between the relative orientations of the two layers. We account for this effect in our CCA approximation through the inclusion of a ``mean-field" term. The result of this term is that the entropy is different for $J_\perp >0$, and $J_\perp < 0$. The expressions for these cases are 
\begin{equation} \label{s_dec_eqs}
   S_\mathrm{dec.} = 
        \cases{
          \log(2) + \varsigma \ \left(1 + \frac{M e^{N J_\perp / T} -  e^{-N J_\perp / T}} {M e^{N J_\perp / T}+  e^{-N J_\perp / T}} \right),\qquad J_\perp >  0          \\
         \log(2) + \varsigma \ \left(1 + \frac{e^{N J_\perp / T} - M e^{-N J_\perp / T}}{e^{N J_\perp / T}+ M e^{-N J_\perp / T}} \right),\qquad  J_\perp <  0   }
\end{equation}
where $M = (1 + m_0^2)/(1-m_0^2)$, with $m_0$ being the magnetisation of the square lattice Ising model ( as given by \Cref{sq_latt_mag} in \ref{app_sq_latt}), and $\varsigma$ is defined as 
\begin{equation}
    \varsigma =  \frac{1}{2}\log\cosh\left(\frac{2 J_\Delta}{T}\right) - \frac{J_\Delta}{T}\tanh\left(\frac{2 J_\Delta}{T}\right).
\end{equation}
Now that we have a way to calculate both $S_\mathrm{b}$ and $S_\mathrm{dec}$ through the Correlated Cluster Approximation, we are able to study the thermodynamics\footnote{The regions of the phase diagram, \Cref{fig:phase_diagram}, that are of interest are in close proximity to first order phase transitions which limits the utility of Monte Carlo simulations. A more advanced numerical study is of course a logical extension to this present work.}. Whilst the CCA provides a framework for the illustrating the thermodynamics of the model predicted by the phase diagram in \Cref{fig:phase_diagram}, it by no means provides quantitively exact results. We aim to go beyond the schematic with this approximation, but the lack of systematic, or analytic, methods available, the approximation is certainly limited. Nevertheless, we find the ``physical" accuracy of the CCA is sufficient for the present discussion -- we give further details in \Cref{app_ICM}. 
We begin the interrogation of the thermodynamics by looking at the first order phase transition through a study of the entropy.

\subsection{First order transition }

The line of first order phase transition is given by $J_\perp(T) = 0$, and hence we are interested in small values of this inter-layer coupling primarily. To understand the behaviour of the entropy, we can consider an expansion of the two terms in \Cref{s_dec_eqs} for small $J_\perp(T)$: taking $e^x \approx 1+x$. Then, taking the difference between these two cases, i.e. across the first order transition, we find that this difference is proportional to $2 m_0^2$. This expansion assumes that we are within a region where the correlation length is small as well. We find that an expansion of the free energy makes the physics more transparent. Indeed, expanding the free energy within our CCA, for the $\pm$ regions of $J_\perp$ around $J_\perp=0$, we find 
\begin{equation}\label{free_energy_exp}
    F_{\mathrm{dec.}} = \cases{
        g - J_\perp m_0^2 + \frac{J_\perp^2 \xi}{2T}(1-m_0^4) + \mathcal{O}(J_\perp^3\xi^2),\qquad J_\perp >0\\
        g + J_\perp m_0^2 + \frac{J_\perp^2 \xi}{2T}(1-m_0^4)+ \mathcal{O}(J_\perp^3\xi^2),\qquad J_\perp < 0  , }
\end{equation}
where $g$ is unimportant for the physics, and vanishes in the thermodynamic limit.

We can see that the two cases in \Cref{free_energy_exp} only differ by the term linear in $J_\perp$. Upon taking the derivative, we obtain $-2m_0^2\ \partial J_\perp / \partial T$, which after computing the derivative for small $J_\perp$, we find that the change in entropy across the first order transition is $\Delta S_\mathrm{dec.} = m_0^2 \ \log(2) $. We have made use of $\partial J_\perp / \partial T|_{J_\perp = 0} = -\log(2) /2$, which is independent of any other parameters. This result is accurate away from the bi-critical point, where the correlation length is small, and contains the only singular contribution. However, there are two important contributions to the entropy below $T_c$; the term linear in $J_\perp$ which we have just discussed, but also there are the non-singular contributions from the quadratic (and higher order) terms, which are important for large $\xi$. 

These higher order terms, whilst continuous away from $T_c$, can provide a rather sharp contribution to the entropy. We see this through the dependence on the correlation length. Far away from the bi-critical point, deep in the ordered phase, the correlation length will be small, and hence the linear term provides the dominant contribution. However, as we approach the bi-critical point, and hence the second order transition, the correlation length increases, as it tends to infinity exactly at the second order transition. The distance between the second and first order transitions decreases as $J_\Delta$ is increased, meaning that the physics close to the second order transition begin to manifest in the entropy. The expansion in \Cref{free_energy_exp} for small $J_\perp$ fails close to the bi-critical point as the result of these higher-order terms become more significant. This necessitates the use of the expression in \Cref{s_dec_eqs}, which is more accurate away from the bi-critical point. 

\begin{figure}
    \centering
    \includegraphics[width=0.8\linewidth]{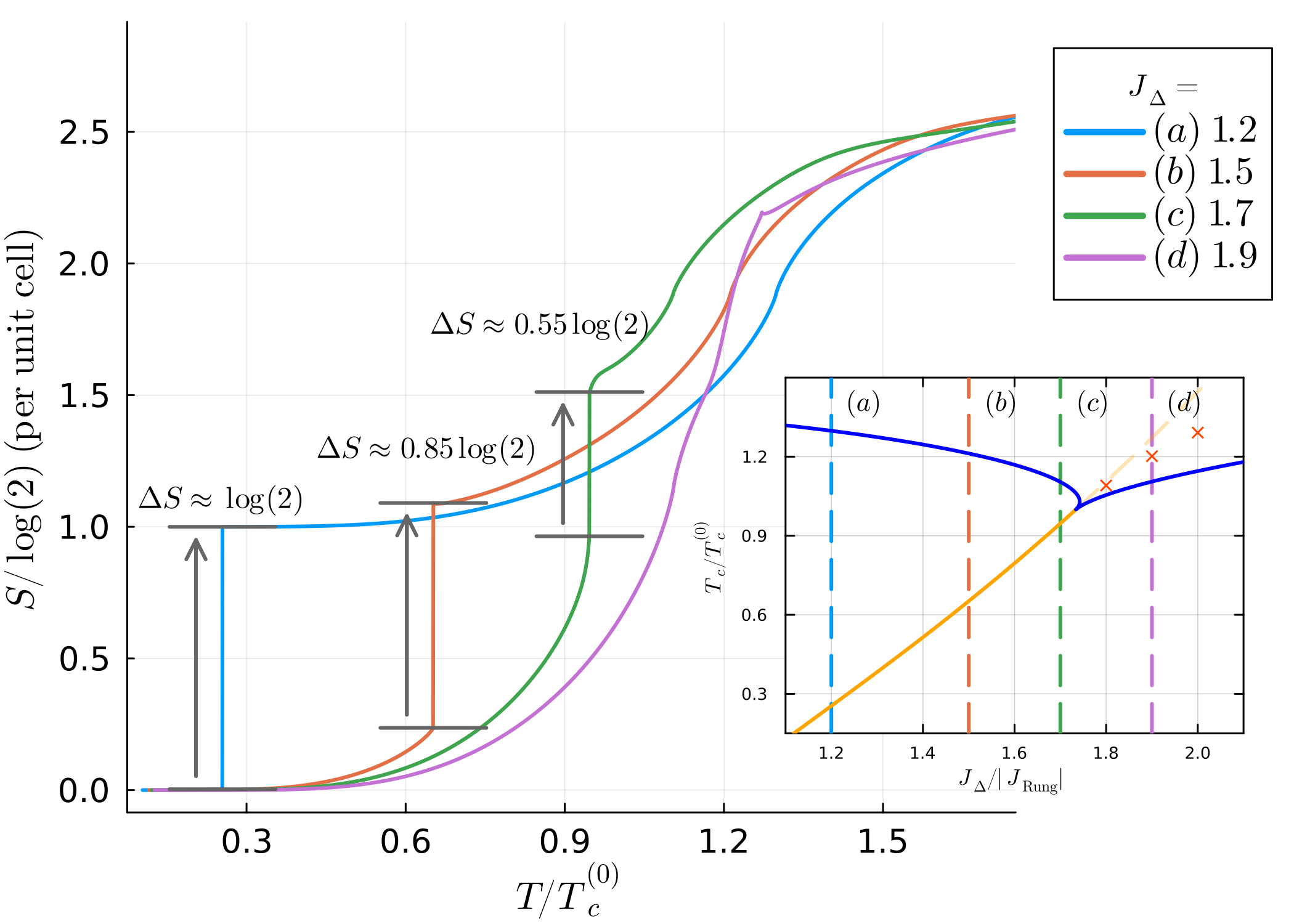}
    \caption{Entropy as a function of temperature for $J_\Delta = (a)\ 1.2, (b)\ 1.5, (c)\ 1.7, (d)\ 1.9$. Details of the approximations used are described in the main text. The discontinuous jump (first order transition) is shown as $\Delta S$. As $J_\Delta$ is increased, the discontinuous jump decreases in size, but we may note that the ``sharply increasing" region contributes more as we increase $J_\Delta$, corresponding to the longer correlation length close to the second order line. The inset shows the corresponding ``scan" through the phase diagram for each of these parameters. This serves as a guide for where one expects to observe the first order transition, for example.}
    \label{fig:entropy_plot}
\end{figure}

We can see this behaviour explicitly in \Cref{fig:entropy_plot}. For $J_\Delta=1.2$, the first order line is crossed at $T \ll T_c$, and we see that the increase in the entropy is entirely described by the linear term with $\Delta S / \log(2) \approx 1$, which agrees with expansion result, with the magnetisation at this temperature $m_0^2 = 1$. The next curve to consider is that with $J_\Delta =1.5$ where we see that the discontinuous increase in the entropy is now $\Delta S /\log(2) \approx 0.85$. The magnetisation in this case is $m_0^2 \approx 0.98$, which is clearly larger than the size of the jump in the entropy. This discrepancy is attributed to the contribution from the quadratic, and higher order terms. We see this as well for for $J_\Delta=1.7$, where the change in entropy is reduced further to $\Delta S /\log(2) \approx 0.55$. Here, the magnetisation is $m_0^2 \approx 0.71$, and we see the largest discrepancy between the expansion and the full expression for the entropy. Nevertheless, this small $J_\perp$ expansion does capture a lot of the essential physics. We now consider the behaviour of the specific heat capacity. 

\subsection{Second order transitions \& the Widom line}

In the same way that we can relate the entropy of the decorated and undecorated models, we can obtain a similar exact relation for the specific heat. There are spurious, non-physical discontinuities/singularities that can arise however. We write the specific heat as
\begin{equation}
    C = 2C_0\left(\frac{T}{T_c} T_c^{(0)}\right) + C_{\mathrm{dec.}},    
\end{equation}
where $C_{\mathrm{dec.}}$ is not exactly equal to the derivative of $S_{\mathrm{dec.}}$, but is chosen to avoid most of the non-physical discontinuities. The term $2C_0$ is obtained by taking the second derivative of the free energy of the square lattice Ising model, \Cref{sq_lattice_free_energy}, and then evaluating this at the scaled temperatures $T_c^{(0)}(T / T_c)$. With this, we are able to study the behaviour as we traverse the phase diagram.

\begin{figure}
    \centering
    \includegraphics[width=0.8\linewidth]{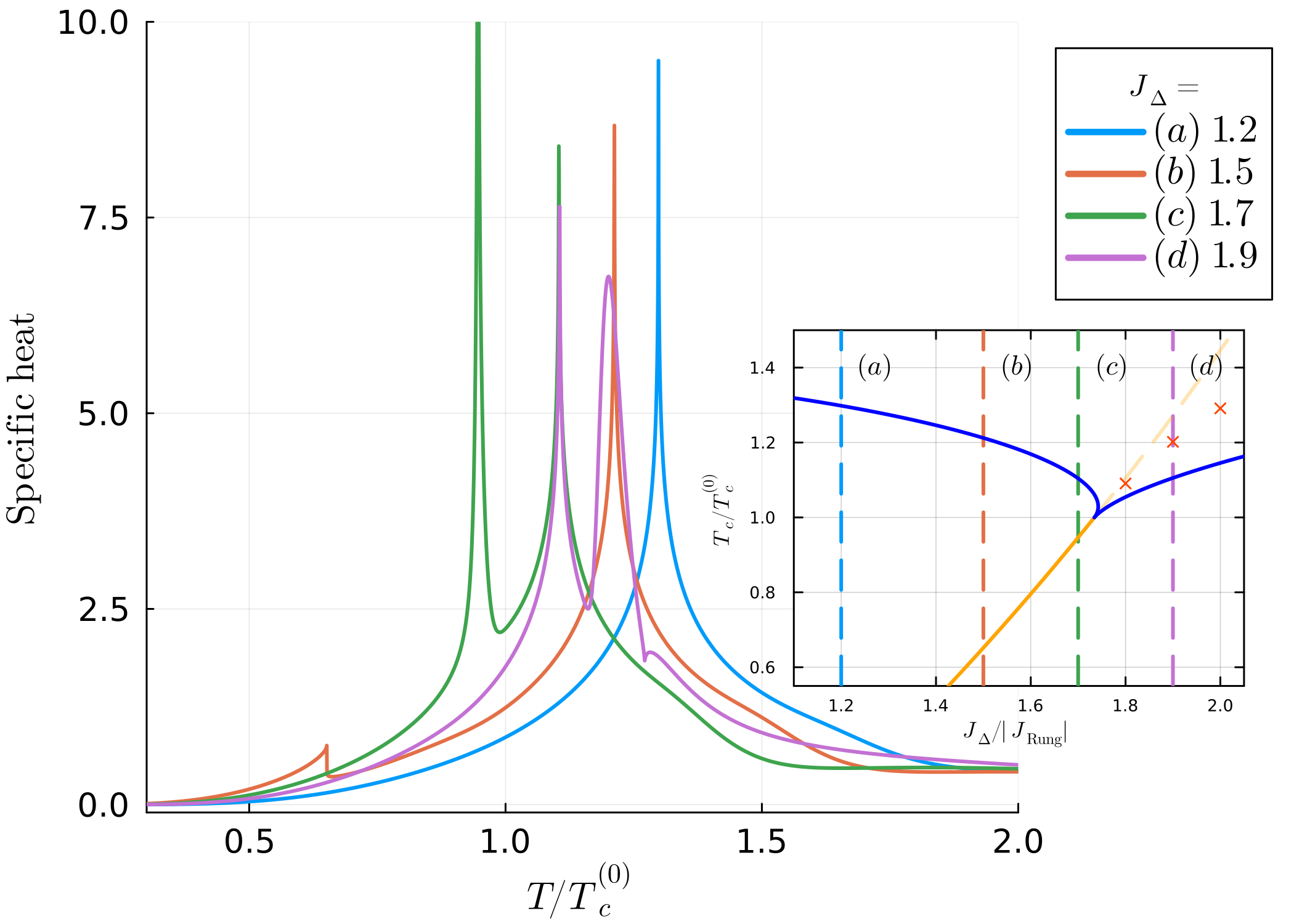}
    \caption{Specific heat for $J_\Delta = (a)\ 1.2, (b)\ 1.5, (c)\ 1.7, (d)\ 1.9$, interrogating three important regions of the phase diagram. These regions are shown in the inset of the figure. The curves for $J_\Delta=1.5,1.9$ have ``kinks" which correspond to the non-analyticity as $J_\perp$ changes sign, and are numerical artefacts in these cases. The change of sign of $J_\perp$ for $J_\Delta = 1.7$ occurs where we see the first peak. This corresponds to the ``pseudo-critical" region close to the second order transition. The inset shows the phase diagram with vertical lines corresponding to the parameters used for the curves in the main figure. This is to aid in identifying the various peaks.}
    \label{fig:spec_heat}
\end{figure}

We begin with the curves shown in \Cref{fig:spec_heat}. The first curve, for $J_\Delta = 1.5$, shows a peak corresponding to the second order transition, but also a low temperature ``kink" which is a numerical artefact of the discontinuity as $J_\perp$ changes sign. The curve for $J_\Delta = 1.7$ features two sharp peaks, one of which can be identified as a second order phase transition, whilst the other occurs where $J_\perp =0$. One might expect that this is another numerical artefact, however, the finite width of this peak leads us to suggest that, in combination with the long correlation length near this point, we are in essence seeing some signs of the Widom line. In a similar vein, we chose $J_\Delta = 1.9$ to more directly probe this Widom line. For this choice of $J_\Delta$, there is no crossing of the first order line, yet we still observe an additional peak in the vicinity of $J_\perp$ changing sign. Indeed, it is in the vicinity of this change of sign that we see this peak; one can observe there is again a ``kink" where $J_\perp$ goes to zero that does not align with the observed peak in the specific heat. This is, in a way, not surprising. Even at the level of the one-dimensional ``Toblerone model" the change of sign of $J_\perp$ did not always align with the observed ``pseudo-transition" - particularly for intermediate or large values of $|J_\Delta - J_R|$ \cite{chapman2024bifurcation}. This is ultimately due to the behaviour of the correlation length as a function of temperature; which through our approximation depends on the singular $T_c$ scaling relation in \Cref{Tc_scaling_mc}.

\begin{figure}[ht]
    \centering
    \includegraphics[width=0.8\linewidth]{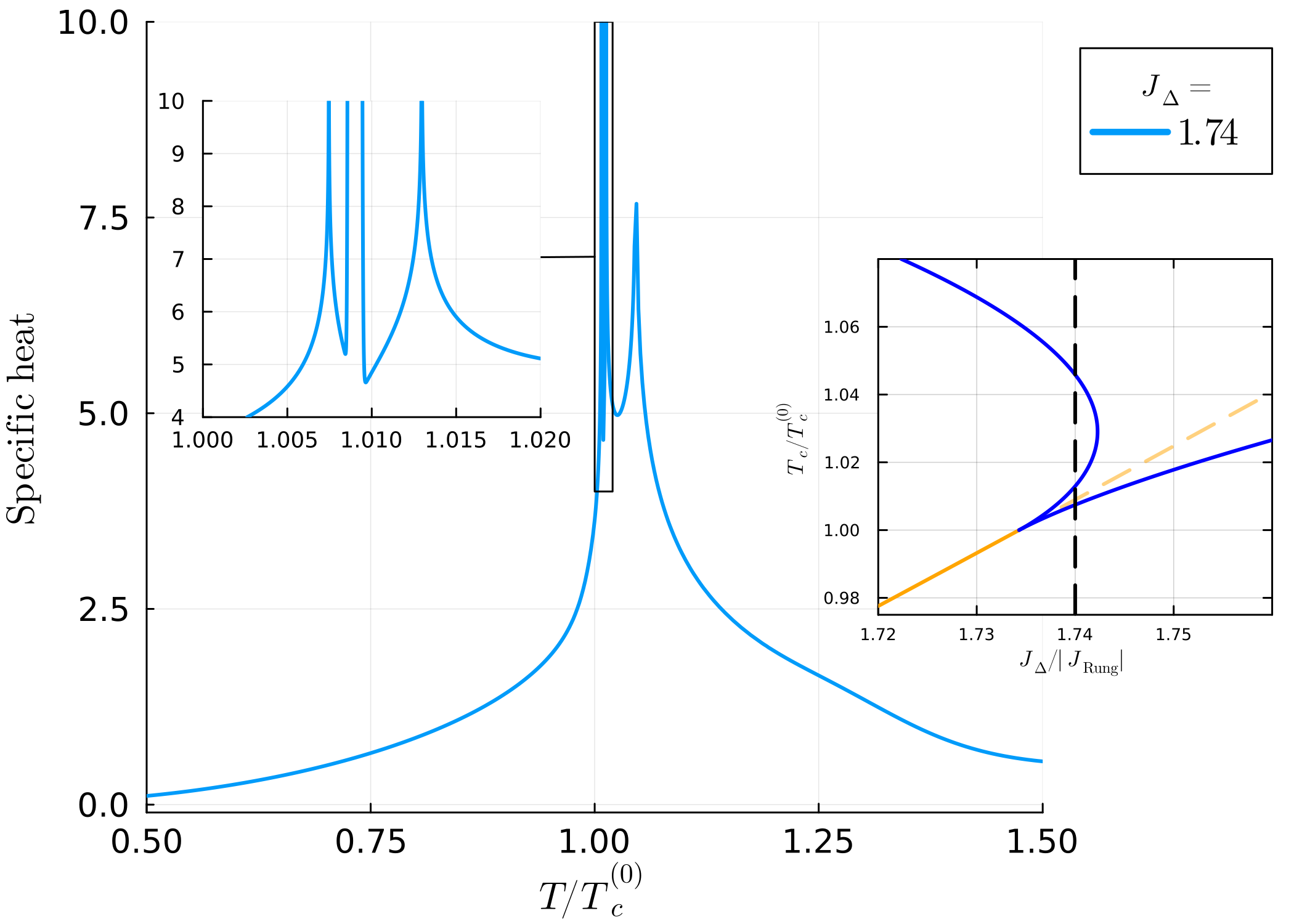}
    \caption{Specific heat for $J_\Delta = 1.74$, which illustrates the re-entrant phase transition, and the Widom line, as predicted by the phase diagram. The left inset shows a zoomed view of three of the peaks ($T/T_c^{(0)} \sim 1$), corresponding to a second order peak, the Widom line crossing, and another second order peak. The other peak in the main figure is also a second order peak. The sequence of peaks is illustrated in the right inset which shows the phase diagram around the region of interest.}
    \label{fig:spec_heat_174}
\end{figure}

This link to the 1D physics will prove insightful shortly, but first, we discuss one more region of the phase diagram: the region in which the second order line curves back on itself, permitting re-entrant transitions. To probe this region, we choose $J_\Delta =1.74$. The plot of the specific heat for $J_\Delta =1.74$ is shown in \Cref{fig:spec_heat_174}. There are four peaks in this specific heat, as predicted from the phase diagram, which we discuss in turn.
The first peak corresponds to crossing the second order line for $J_\perp >0$. This is almost immediately followed by another, sharp peak as we cross the Widom line in the disordered phase. Upon increasing temperature further, we become ordered again, re-entering the AFM phase. The final peak corresponds to exiting this phase and entering the paramagnetic phase once more. This complex landscape of peaks arises from the singular, slower-than-linear, dependence of $T_c$ on $J_\perp$. What's more, we can begin to appreciate the signature of the Widom line more. The coincidence of the Widom line with the change of sign of $J_\perp(T)$ brings us to our final point of discussion in the following section. 

\section{The Widom Line Interpretation}\label{sec_widom}

We now return to the Widom line in the phase diagram. In \Cref{fig:spec_heat_174} (and curve $(d)$ in \Cref{fig:spec_heat}), it can be observed that the specific heat exhibits an additional peak that corresponds to crossing the Widom line. Indeed, this peak can be distinguished from those corresponding to second order transitions. The Widom line peak is finite, and broader than those peaks corresponding to the second order transitions. 

Indeed, one can observe the similarity between these Widom line peaks in the two-dimensional model, and the observed pseudo-transitions in the one-dimensional models. Both feature sharp, but finite peaks in the specific heat, however in the phase diagram of the higher dimensional model, the line of these peaks extend from the critical endpoint of a line of first order transitions. Viewing these phenomena through the lens of some higher dimensional model (where phase transitions can occur at finite temperature) provides deeper, physical insights into the pseudo-critical behaviour in 1D. If one could ``extend" the $T=0$ region of the one-dimensional model, the present work suggests that a similar structure might exist to the phase diagram in \Cref{fig:phase_diagram}, where the ``pseudo-transition peaks" could be recognised as extending from the end point of a line of first order transitions. Indeed, this first order transition between two phases, with distinct entropies, is a necessity for the occurrence of the pseudo-transition in the corresponding one-dimensional model.

The slope of this first order line can be understood through the Clausius-Clapeyron relation \cite{carnot1872reflexions,reichl2016modern}, where the line slopes towards the lower entropy phase. This is clearly seen in \Cref{fig:phase_diagram}, where the line slopes towards the ferromagnetic phase, which has a constant entropy. Further, with the distinct entropies of these neighbouring phases, we may observe the first order transition, and hence the Widom line, as a function of temperature. 
We can also consider the limit where these two phases have the same entropy as $T \to 0$. Here, the first order line becomes vertical at low temperature, but may still slope at intermediate temperature. This has the consequence that the Widom line will also remain sloped, and one may still observe the crossing as a function of temperature. Indeed, we suggest that this will persist to the corresponding one-dimensional model, where it is expected that a pseudo-transition will be observed.

This discussion is aimed towards reconciling this physical argument in terms of the first order line and Widom line, with the mathematical arguments of Ref.~\cite{rojas2020conjecture}. The cases we have discussed so far are in good agreement with those in Ref.~\cite{rojas2020conjecture}, which play host to pseudo-transitions at finite temperature. Through this new lens of the Widom line, we obtain a deeper physical insight into the origin of these pseudo-transitions in one-dimensional models. 

We also briefly address the two cases in Ref.~\cite{rojas2020conjecture} which do not show evidence of a pseudo-transition. In these two cases, the entropy of two neighbouring phases is no longer continuous at the phase boundary - the ``critical residual entropy" is not equal to that of either phase. The interpretation in terms of a Widom line here is less obvious. We might try to understand these cases by considering what the phase diagram might look like in higher dimensions. However, the construction of a model that has the required phase boundaries is not clear at present, and falls beyond the scope of this work. We do offer the comment that the discontinuity of this critical entropy (i.e. where this entropy differs from that of the neighbouring phases) may suggest an additional phase in the vicinity. If this is indeed the case, there are now three phases that meet around this first order transition. We leave further investigation of this to future work.

Whilst the interpretation of the pseudo-transitions in terms of a Widom line that we present here is largely phenomenological, we have shown that there are deeper insights, and further utility that can be obtained from this. Indeed, the existence of the Widom line in the 2D model is quite an interesting phenomena in itself, adding to the rich phase diagram in \Cref{fig:phase_diagram}.

\section{Conclusions}

In this work we have studied a two-dimensional decorated bilayer Ising model, obtaining both the phase diagram and plots of the thermodynamic quantities. The phase diagram is obtained through a semi-empirical approach in which we fit a scaling relation to unbiased Monte-Carlo data. To complement this, we study the thermodynamics of the model through the introduction of a novel, physically motivated approximation method, which is found to capture the essential physics in a one-dimensional Ising model (for which there is an exact solution to compare to). Indeed, the comparison to the one-dimensional Ising model is a direct motivation this work. The model studied in this present work was a direct generalisation of the previously studied ``Toblerone model" that was found to exhibit a ``pseudo-transition" \cite{yin2023marginal,hutak2021low,chapman2024bifurcation}.

Through this work, we have uncovered a collection of deeper insights into the physics of the decorated Ising models. The singular scaling of the critical temperature gives rise to re-entrant phase transitions. We also find a bi-critical point where we derive the new critical exponents for the enhanced symmetry at this point. This bi-critical point occurs at the critical end-point of a line of first order transitions. Extending from this point, we observe sharp, but finite, peaks in the specific heat which we interpret as a \textit{Widom line}. This Widom line has implications for the understanding of the physics of the aforementioned one-dimensional models. We have demonstrated here that the physics of the ``pseudo-transitions" observed in these decorated models can be viewed as crossing a Widom line extending from a critical end-point at zero temperature. 

The relation to the Widom line provides new physical grounding for the phenomenon of the ``pseudo-transition", and indeed existence of the same. Prior to this, the mathematical arguments of Rojas in Ref.~\cite{rojas2020conjecture} provided an existence criterion for the occurrence of the pseudo-transition. However, we now understand that the structure of the phase diagram in some higher-dimensional model can arrive at the same existence conditions. The other cases in Ref.~\cite{rojas2020conjecture}, where the residual entropy is no longer continuous, do not have a clear interpretation in terms of the  phase diagram of some higher-dimensional model. Understanding these phase boundaries is certainly an compelling question. It will indeed be interesting to extend these findings to the other models that have been found to undergo the pseudo-transition. 

\ack
JC was supported by the Engineering and Physical Sciences Research Council [EPSRC DTP 2022 EP/W52461X/1]. BT and JC acknowledge partial funding support for a visit to the Institut Laue–Langevin with a QuantEmX grant from ICAM and the Gordon and Betty Moore Foundation through Grant GBMF5305. The authors are grateful to Tim Ziman for insightful discussions relating to this work.

\vspace{1em}

\appendix
\section{Summary of results for the square lattice Ising model}\label{app_sq_latt}
The solution of the Ising model on the square lattice, by Onsager \cite{onsager1944crystal}, is still widely regarded as a \textit{tour de force} of statistical physics. The evaluation of the partition function, and exact expression of the free energy was pivotal for the development of the study of phase transitions. In this section, we provide a summary of these results that are relevant to this present study.

The solution of Onsager is indeed infamous, both for its significance and for its complexity. Since then, there have been alternative approaches which provide a more accessible solution, such as those by Kac and Ward \cite{kac1952combinatorial}, and Vdovichenko \cite{vdovichenko1965calculation}. 
The evaluation of the partition function permits us to obtain an expression for the free energy. For the square lattice Ising model, the free energy (per site), $F_0$, can be written as
\begin{equation}\label{sq_lattice_free_energy}
    \eqalign{F_0 = -T\log(2) - \frac{T}{8\pi^2}\int_0^{2\pi}dp\int_0^{2\pi}dq &\ \log[(1+x^2)^2 - \cr  & 2x(1-x^2) (\cos(p) + \cos(q))],}
\end{equation} 
where $x = \tanh(J / T)$. From the free energy, one can in principle obtain any thermodynamic quantity of interest through the appropriate derivatives. First, the entropy can be obtained from the first temperature derivative of the free energy
\begin{equation}\label{sq_latt_entropy}
    \fl \eqalign{S_0 = & \log(2) +  \frac{1}{8\pi^2}\int_0^{2\pi}dp\int_0^{2\pi}dq \ \log[(1+x^2)^2 -  2x(1-x^2) (\cos(p) + \cos(q))] \cr \cr  &+ J\ \frac{1-x^2}{8\pi T^2} \int_0^{2\pi}dp\int_0^{2\pi}dq\  \frac{4x(1-x^2) - 2(1-3x^2)  (\cos(p) + \cos(q))}{(1+x^2)^2 -  2x(1-x^2) (\cos(p) + \cos(q))}.}
\end{equation}

The specific heat is obtained in a similar fashion by taking $C_0 = T d^2F_0 /dT^2$, however the expression gets rather cumbersome and so we omit it here. From \Cref{sq_latt_entropy}, we are able to deduce some important results about the phase transition in the 2D Ising model. We know that a phase transition must coincide with some singularity in the thermodynamics. The singular part of \Cref{sq_latt_entropy} is clearly in the logarithmic term. Taking $\cos(p) + \cos(q) = 2$ yields the critical value of $x$ to be 
\begin{equation}
    \tanh(J / T_c^{(0)}) = \sqrt{2} - 1,
\end{equation}
From this, one finds the critical temperature, $T_c^{(0)}$ to be 
\begin{equation}\label{sq_latt_tc}
    T_c^{(0)} = \frac{2J}{\log(1 + \sqrt{2})} \approx 2.2691 \dots,
\end{equation}
which is the familiar result. Indeed we see that the square lattice Ising model has a spontaneous phase transition at the critical temperature $T_c^{(0)}$. We can obtain details of the behaviour of the specific heat in the vicinity of $T_c^{(0)}$ by introducing the variable $t = T- T_c^{(0)}$, and expanding the entropy for small $t$ (and for $p,q$). Making this expansion and integrating, we obtain
\begin{equation}
    S_0 \sim a\ t \log|t| + \mathrm{non-singular\ terms},
\end{equation}
where $a$ is a constant. Then, taking the derivative, we find that
\begin{equation}
    C_0 \sim a\log |t| + \mathrm{non-singular\ terms}.
\end{equation}
Thus we can see that the specific heat diverges logarithmically as $T \to T_c^{(0)}$. The corresponding critical exponent $\alpha =0$. 
Within this ordered phase, the order parameter is non-zero. For the case of the square lattice Ising model, the order parameter is the magnetisation, which is defined as 
\begin{equation}\label{sq_latt_mag}
    m_0 = \left[1 - (\sinh(2J / T))^{-4}\right]^{1/8},
\end{equation}
which is zero above $T_c^{(0)}$.

Also of interest is the behaviour of the correlation length in the vicinity of the critical point. At a phase transition, fluctuations on all length scales become important. For the square lattice Ising model, the correlation length follows the scaling relation 
\begin{equation}\label{sq_latt_xi}
    \xi \sim |T - T_c^{(0)}|^{-\nu},
\end{equation}
where the critical exponent $\nu = 1$, and this relation is valid near the critical point. One observes that the correlation length also diverges as we approach $T_c^{(0)}$.

\section{Monte Carlo Results}\label{app_MC}

\begin{figure}
    \centering
    \includegraphics[width=0.8\linewidth]{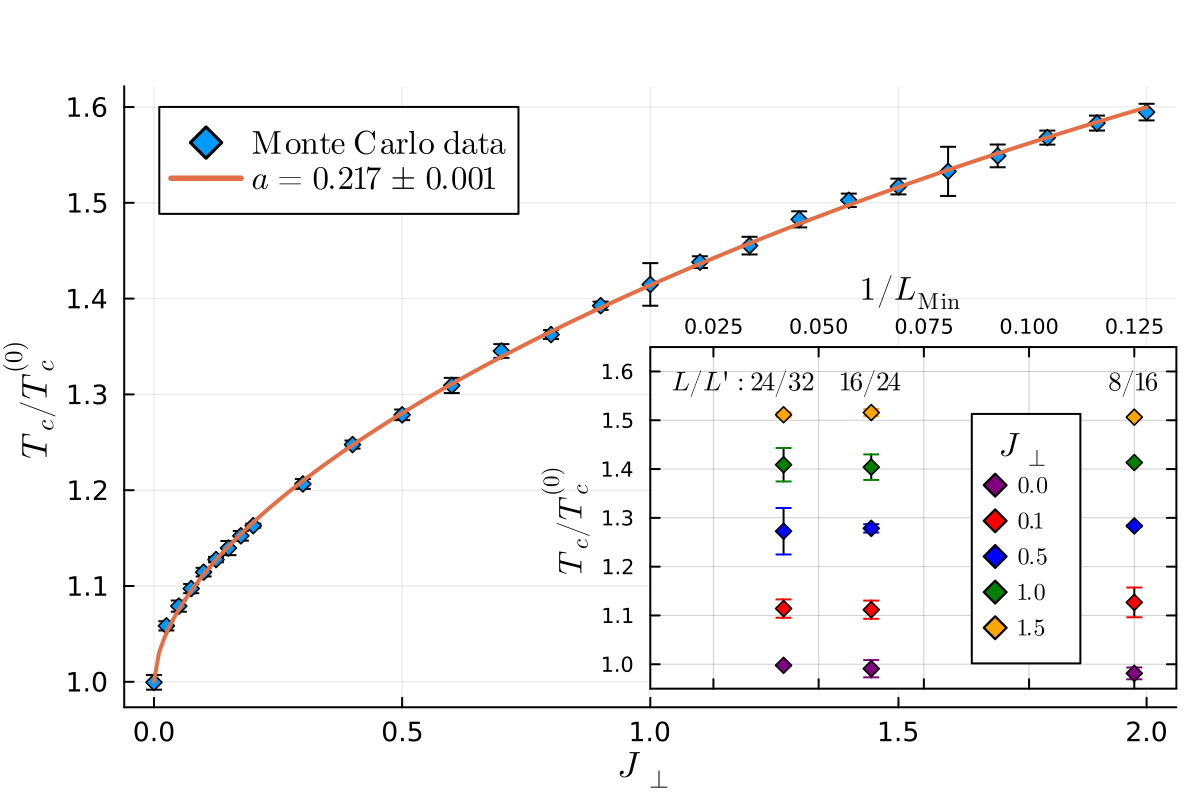}
    \caption{Fitting the interpolation function, \Cref{eq:interpol_tc}, to critical temperature data from Monte Carlo simulations. The critical temperature is obtained by locating the crossing point for Binder Cumulants, \Cref{binder_eq}, at two distinct system sizes $U_4(L)=U_4(L')$. The inset shows that this estimate does not vary significantly with system size, where we plot the critical temperatures for $J_\perp = 0,0.1,0.5,1.0,1.5$ for Binder cumulants with pairs of $L$ values $(L,L') = (8,16), (16,24), (24,32)$. Error bars are obtained through a jackknife resampling of the data. }
    \label{fig:mc_fit_tc}
\end{figure}

In order to deduce an expression for the critical temperature scaling that is valid for a wider range of $J_\perp$, we make use of the unbiased numerical results from Monte Carlo simulations. The focus of this paper is not on advanced numerics, and so the details of the MC implementation are kept brief. The Wolff cluster algorithm \cite{wolff1989collective} is implemented in the Julia programming language. Lattice sizes of $N = 16\times16\times2$, and $N =32\times32\times2$ are used, with $7.5\times10^4$ update sweeps, with measurements taken every 100 steps. 
To estimate the critical temperature for each value of $J_\perp$, the Binder Cumulant, $U_4$ is calculated, as given by
\begin{equation}\label{binder_eq}
    U_4 = 1 - \frac{\langle m^4 \rangle}{3 \langle m^2 \rangle^2}.
\end{equation}
From \Cref{binder_eq}, we may obtain the critical temperature by finding the intersection of the curves of $U_4$ for successive values of the system size \cite{binder1992monte}. 
The critical temperature is extracted for a range of values for $ 0 < J_\perp \leq 2$. This allows fitting to the semi-empirical relation
\begin{equation}\label{eq:interpol_tc}
    T_c = T_c^{(0)}\left(1 + |\tanh(aJ_\perp)|^{4/7}\right),
\end{equation}
where $a$ is left as a fitting parameter. From our simulation, we find that $a = 0.217 \pm 0.001$, but we stress here that the precise values is unimportant to the physics of the model. The features of the phase diagram do not depend on the value of $a$. This fit is shown in \Cref{fig:mc_fit_tc}, where the error bars on each $T_c$ estimate are obtained through a jackknife resampling of the data. By way of benchmarking, this implementation gives a good estimate for the known critical temperature for the square lattice Ising model when $J_\perp =0$: $T_c^{(0)} \approx 2.267\pm0.03$. 

Determining the critical temperature through the intersection of the Binder cumulants is insensitive to the system size, as shown in the inset of \Cref{fig:mc_fit_tc}. Indeed, we do not require high numerical precision for the present purposes, but one observes that the scaling relation \Cref{eq:interpol_tc} fits well to the Monte Carlo data.

\section{Critical Temperature and exponent derivations}\label{app_tc_exp}

\subsection{Critical temperature scaling}\label{app_tc_scaling}
To begin, it is noted that in some ordered phase, the two layers exert a magnetic field on each other, given by $B=|J_\perp| m $, where $m$ is the magnetisation of one of the layers. Using the known scaling behaviour of the magnetisation with the field at the unperturbed critical temperature $T_c^{(0)}$, $m\sim B^{1/\delta}$, we find (the relevant critical exponents for the 2D Ising model are given in \Cref{tab:Exponents})
\begin{equation}
    m\sim |J_\perp| ^{1/(\delta -1)}.
\end{equation}
This means that at $ T_c^{(0)}$ there is a non-zero spontaneous magnetisation which implies that the critical temperature has been increased due to the inter-layer interaction. We can see how much this increase is by using the scaling relation of the magnetisation with temperature, $m\sim (T_c - T)^\beta$. This gives us the relation between the increase in transition temperature and the inter-layer coupling as
\begin{equation}
    T_c -  T_c^{(0)} \sim |J_\perp|^{1/\beta(\delta -1)}.
\end{equation}
The exponent of $J_\perp$ can be rewritten through Widom's exponent relation $\beta (\delta -1)= \gamma $ \cite{stanley1971introduction,cardy1996scaling}, we obtain the result quoted in the main text
\begin{equation}\label{app_tc_jperp}
    T_c - T_c^{(0)} \sim |J_\perp|^{1/\gamma}.
\end{equation}

While this derivation may look like it is based on a mean-field argument for the coupling between the layers, the use of the known critical exponents means the result is exact.  This can be independently verified by a completely different method, based on the mapping between the two-dimensional classical Ising model and the one-dimensional quantum Ising model in the vicinity of the critical point \cite{gogolin2004bosonization}.  We sketch the derivation here without details as it is not important for the other results.

The two coupled Ising planes are first converted to two coupled quantum Ising chains, then rewritten as a bosonic field theory with Hamiltonian \cite{fabrizio2000critical}
\begin{equation}
\label{eq:DSG}
    H = \frac{v_0}{2} \left[ \Pi(x)^2 + (\partial_x \phi)^2 \right] - g \cos \beta\phi(x) - \lambda \sin[(\beta/2)\phi(x)].
\end{equation}
Here, $v_0$ is a Fermi-velocity that sets the energy scale, $g \propto (T-T_c^{(0)})$, and $\lambda \propto J_\perp$.  The parameter $\beta=\sqrt{4\pi}$ for an Ising model — a more general model where $\beta$ is not fixed was discussed in Ref.\cite{fabrizio2000critical}.  The original spins can be related to exponents of the field $\phi$ \cite{gogolin2004bosonization}, and $\Pi(x)$ is the conjugate momentum to the field $\phi(x)$.

This Hamiltonian is known as a double frequency sine-Gordon model.  Each perturbation alone would open a gap in the spectrum, with size $m_g \sim g^{\frac{1}{2-\beta^2/4\pi}}$ and $m_\lambda \sim \lambda^{\frac{1}{2-\beta^2/16\pi}}$.  The transition occurs when the two masses are equal , i.e.
$\lambda_c(g) \sim g^\gamma$ where \footnote{The expression for $\gamma$ in section 2 of Ref.~\cite{fabrizio2000critical}  (where it is called $\nu$) has a typo in it — the correct expression is given here.  We thank the authors of \cite{fabrizio2000critical} for bringing our attention to this correction.}
\begin{equation*}
\gamma = \frac{32\pi - \beta^2}{32\pi - 4\beta^2}.
\end{equation*}
Putting in $\beta^2=4\pi$ for the Ising model gives us $\gamma=7/4$ and hence reproduces \Cref{app_tc_jperp} above.

\subsection{Bi-critical point exponents}\label{app_exponents}

In general, the bilayer Ising model is in the same universality class as the standard square lattice Ising model. Indeed, in Ref.~\cite{abe1970some}, the bilayer is classified based on the unchanged critical exponents. As the inter-layer coupling goes to zero at the bi-critical point, there is a change to the critical exponents. These exponents are now derived. We begin from the scaling of the critical temperature
\begin{equation}\label{app_tc_eq}
    T_c = T_c^{(0)} (1 + a|J_\perp|^{4/7}).
\end{equation}
We note here that close to the bi-critical point $J_\perp \propto -t $ where $t = T- T_c^{(0)}$. We can then use this result for the undecorated bilayer which should obey the typical Ising scaling relations. Namely, we can begin with the magnetisation scaling relation
\begin{equation}\label{mag_scaling}
    m \sim (T_c - T)^{1/8}. 
\end{equation}
The substitution of \Cref{app_tc_eq} into the scaling relation \Cref{mag_scaling} then yields 
\begin{equation}
    m \sim |(T-T_c^{(0)}) - a(T-T_c^{(0)})^{4/7}|^{1/8}.
\end{equation}
 In the vicinity of the bi-critical point, the term with the $4/7$ power will dominate over the linear term, which can then be neglected. This then gives
 \begin{equation}
    m_{\mathrm{BCP}}\sim |t|^{1/14},
\end{equation}
where we can find $\beta = 1/14$ at the bi-critical point. In a similar way, the scaling relation for the magnetic susceptibility is 
\begin{equation}
    \chi \sim (T - T_c)^{7/4} = (t - a|t|^{4/7})^{7/4},
\end{equation}
where again the $4/7$ power dominates over the linear term. This gives the new exponent relation as
\begin{equation}
    \chi_{\mathrm{BCP}} \sim |t|.
\end{equation}
Thus our new exponent at the bi-critical point is $\gamma = 1$. We continue in this fashion and to obtain the full set of exponents as given in \Cref{tab:Exponents}. We find that this new set of exponents satisfies the various exponent relations \cite{stanley1971introduction,cardy1996scaling}
\begin{equation}
\eqalign{
    \beta (\delta -1)= \gamma, \\
    \alpha + 2\beta + \gamma =2, \\
    2 - \alpha = \beta (\delta + 1).  }  
\end{equation}

\section{Formulation of the ``Correlated Cluster Approximation"}\label{app_ICM}
In the main text, we showed in \Cref{entropy_relation} that the entropy of the model \Cref{model_hamiltonian} could be separated into two distinct contributions: the entropy of the bilayer $S_b$, and the contribution from the decorating spins $S_{\mathrm{dec.}}$. The bilayer contribution is responsible for the second order transitions in the model, and the contribution from the decoration gives rise to the first order transitions, and Widom line in \Cref{fig:phase_diagram}.
The entropy of the decoration can be expressed as
\begin{equation}\label{app_ent_dec}
    S_\mathrm{dec.} = \left(1 + \frac{\partial F_b}{\partial J_\perp}\right)\frac{\partial J_\perp}{\partial T} + \log 2.
\end{equation}
The derivative of the interlayer coupling with respect to temperature can be calculated exactly from \Cref{RG_eq}, thus leaving the derivative of the free energy of the bilayer, $F_b$, with respect to $J_\perp$. We were able to obtain an approximation for the entropy of the bilayer in \Cref{S_bil_approx}, by using the exact result for the square lattice Ising model and scaling the temperature. This gives the correct logarithmic divergence in the specific heat upon differentiation. An analogous approach to the calculating the free energy, writing $F_b = 2F_0(T T_c^{(0)}/T_c)$, is not as useful. The anomalous exponent in the critical temperature as a function of $J_\perp$, \cref{Tc_scaling_mc}, gives spurious and unphysical divergences at $J_\perp=0$ when put into \Cref{app_ent_dec}. This motivates the search for an alternative method for the calculation of $F_b$. 

One might consider a formal expansion of $F_b$ in the interlayer coupling $J_\perp$; the physics of interest in \Cref{fig:phase_diagram} occurs for small $J_\perp$. Following Ref.~\cite{abe1970some}, an expansion to second order gives the dominant behaviour as
\begin{equation}
    F_b \approx 2F_0 - |J_\perp| m_0^2 + (J_\perp^2/T) \sum_{i,j}\ [ \langle s_i s_j \rangle ^2 - \langle s_i \rangle^2 \langle s_j \rangle^2],
\end{equation}
where the summation is over spins in a single layer. The summation cannot be evaluated exactly, however we can note that the summand is approximately a correlation function, allowing us to approximate this by $\exp(-r/\xi)$. We then obtain a term proportional to $\xi^2$, and write the expansion as
\begin{equation}\label{f_expansion}
    F_b \approx 2F_0 - |J_\perp| m_0^2 + (J_\perp \xi)^2 (2 m_0^4 /T) + \dots,
\end{equation}
where $2F_0$ is the zeroth order term in $J_\perp$ from the two uncoupled square lattice Ising models, and all other symbols have their previously defined meanings. This expansion is by no means exact, but gives an idea of the significant contributions. We see that the term linear in $J_\perp$ can be related to the first order transition, giving the jump in the entropy, proportional to the magnetisation. However, by the second order term, we see that this expansion is no longer in terms of solely $J_\perp$, but in terms of $(J_\perp \xi)^2$. This is clearly not small in the vicinity of the second order transition, meaning the expansion is no longer well controlled. This is certainly an issue as the interesting physics in \Cref{fig:phase_diagram} is around $J_\perp=0$, for which this second order term may still be large. The lack of a systematic expansion is a driving motivation to introduce an alternative way to capture the physics of the model.

\begin{figure}[htbp]
    \centering
    \begin{subfigure}[b]{0.47\textwidth}
        \includegraphics[width=\linewidth]{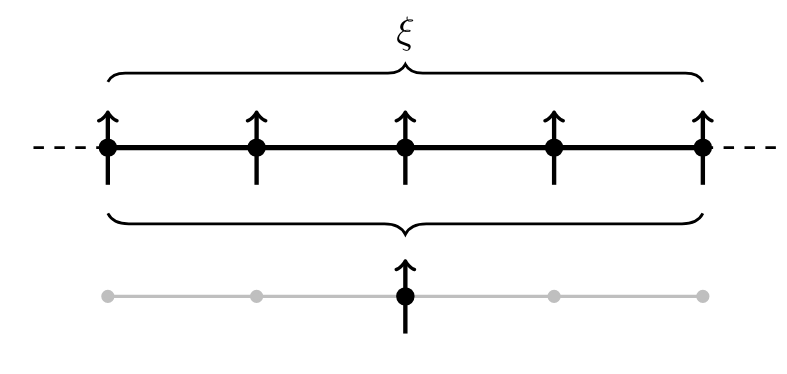}
        \caption{The first step of correlated cluster approximation, in which the spins on each leg of the ladder can be reduced to one \textit{effective} spin which represents the spin state within the correlation length. }   
    \end{subfigure}
    \hfill
    \begin{subfigure}[b]{0.47\textwidth}
        \includegraphics[width=\linewidth]{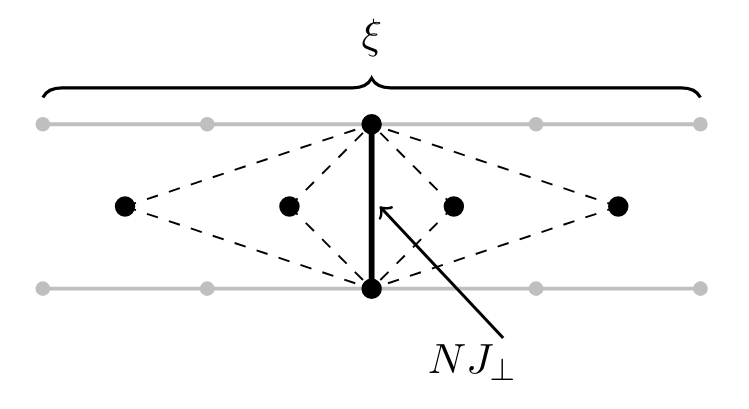}
        \caption{After the initial reduction in (a), we see the resultant lattice, with the single effective dimer coupled to each of the decorating spins. The coupling of the dimer $NJ_\perp$ arises as we have combined $N$ spins.}     
    \end{subfigure}
    \caption{The reduction of a block of spins of size $\xi$ to a single spin for the ``Correlated Cluster Approximation". The spins in grey are shown as a reference to the initial lattice.}
    \label{fig:ICM_lattice_sketch}
\end{figure}

For the purposes of this present work, the physics of interest occurs mostly while the correlation length is long, as was found in Ref.~\cite{chapman2024bifurcation}. Thus the spins within the correlation length are providing the dominant contribution to the thermodynamics of the model. Based on this argument we can begin to formulate our main approximation, which comes when computing the partition function. We choose to restrict the summation to only include states where $s_i^{(1)} = s^{(1)}_{j},\ s_i^{(2)} = s^{(2)}_{j}$ - those that are within the correlation length by definition. This length defines the size of the system, which we then call the ``Correlated Cluster Approximation", or CCA. This permits each of the spins within the correlation length to be reduced to one ``effective" spin. After this, we obtain a single dimer of these effective spins, which are still coupled to the decorating spins. This is sketched for the one-dimensional ``Toblerone lattice" in \Cref{fig:ICM_lattice_sketch}. The procedure naturally extends to two dimensions. 

Now that we have an overview of the approximation, we can begin to obtain thermodynamic quantities. We write the Hamiltonian of this model as 
\begin{equation}\label{ICM_hamiltonian}
    \mathcal{H}_{\mathrm{dec.}} = - \sum J_\Delta(s_i^{(d)})(s_i^{(1)} + s_i^{(2)}) + J_\perp(s_i^{(1)} s_i^{(2)}) + B (s_i^{(1)} + s_i^{(2)}),
\end{equation}
where $s_{i}^{(1)},s_{i}^{(2)},s_{i}^{(d)}$ are the spins on layers 1, 2, and the decorating spins respectively, and $B$ is and effective field that includes any field experienced by spin $s_i^{(1)}$ due to a magnetisation on $s_i^{(2)}$. By restricting the summation to only those states with $s_i^{(1)} = s^{(1)}_{j},\ s_i^{(2)} = s^{(2)}_{j}$ (i.e. all $s^{(1)}_i$ are the same, and similarly for $s_i^{(2)}$), we obtain the partition function as 
\begin{equation}
        Z_\mathrm{dec.} = 2^{N+1}\left[e^{-NJ_\perp} + e^{NJ_\perp}\cosh(2NB)\cosh^N(2J_\Delta)\right]. 
\end{equation}
The $N$-dependence has been left explicit so that we may highlight that for this approximation. For the 2D model in the main text, we take $N = \xi^2$, with $\xi$ given in \cref{sq_latt_xi}. Then, the free energy is then obtained in the usual way, $F_{\mathrm{dec.}}~=~-~T~\log(Z_\mathrm{dec.})$, whereupon differentiation, can be used to give \Cref{s_dec_eqs} in the main text. 
As mentioned previously, the field $B$ accounts for the effects of a magnetisation on one of the spins $s_i^{(1,2)}$ on the other $s_i^{(2,1)}$. Thus it is natural to express this field in terms of a magnetisation. We can obtain such an expression by first expanding for small $J_\perp$, and then finding the magnetisation within this ``Correlated Cluster Approximation". Then, for $J_\perp =0$, we invert the magnetisation, and find an expression for the field
\begin{equation}
    B = T \tanh^{-1}(m_0),
\end{equation}
where $m_0$ is the magnetisation when $J_\perp =0$. For the bilayer model considered in the main text, this is the magnetisation of the square lattice Ising model, as given in \Cref{sq_latt_mag}. This result also gives the linear term in the expansion of the free energy as in \Cref{free_energy_exp}, which gives rise to the jump in the entropy across the first order transition. Thus we can be confident that any corrections for small $J_\perp$ are unimportant to the physics. 

\begin{figure}[htbp]
    \centering
    \begin{subfigure}[b]{0.47\textwidth}
        \includegraphics[width=\linewidth]{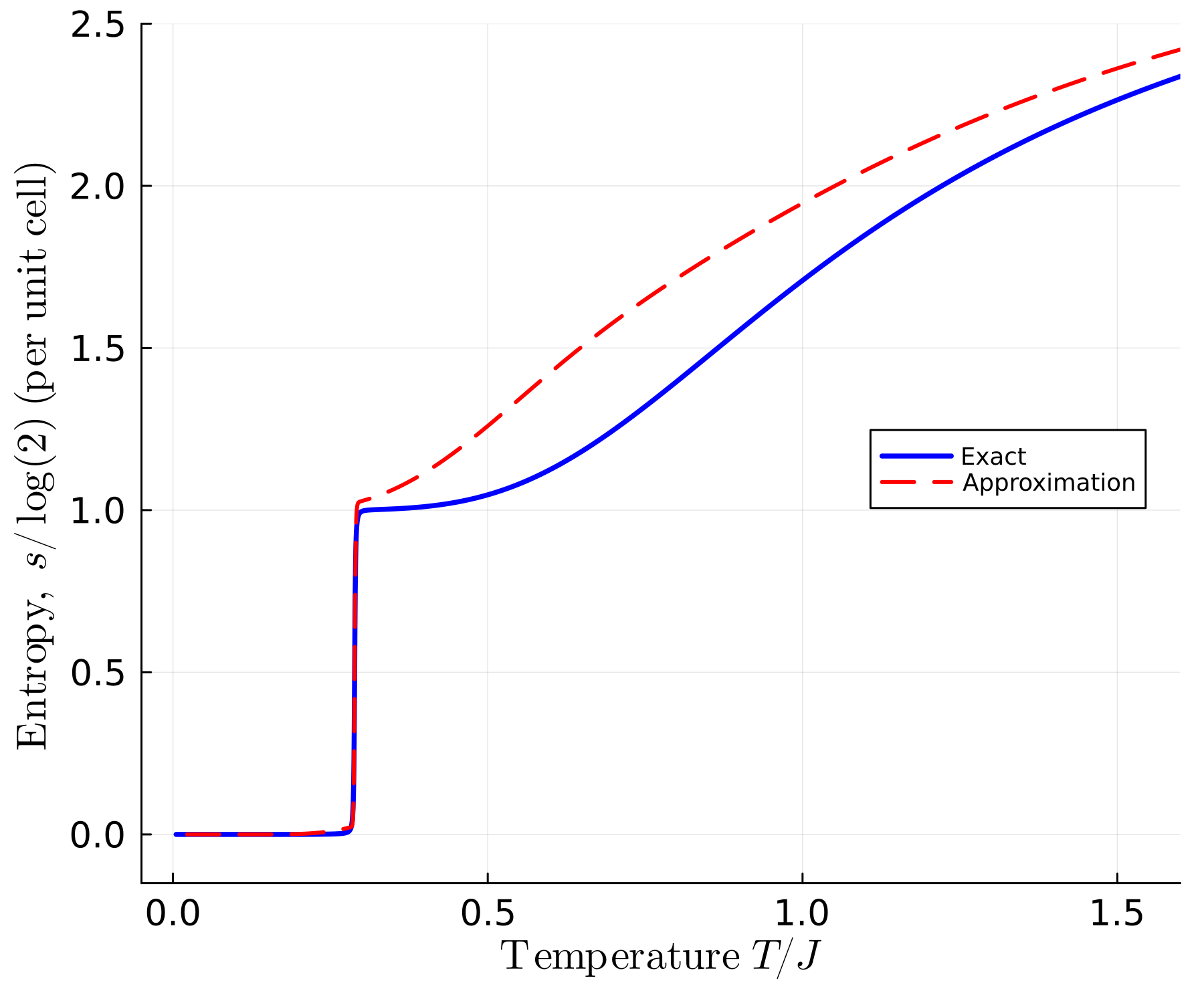}
        \caption{Entropy}     
    \end{subfigure}
    \begin{subfigure}[b]{0.47\textwidth}
        \includegraphics[width=\linewidth]{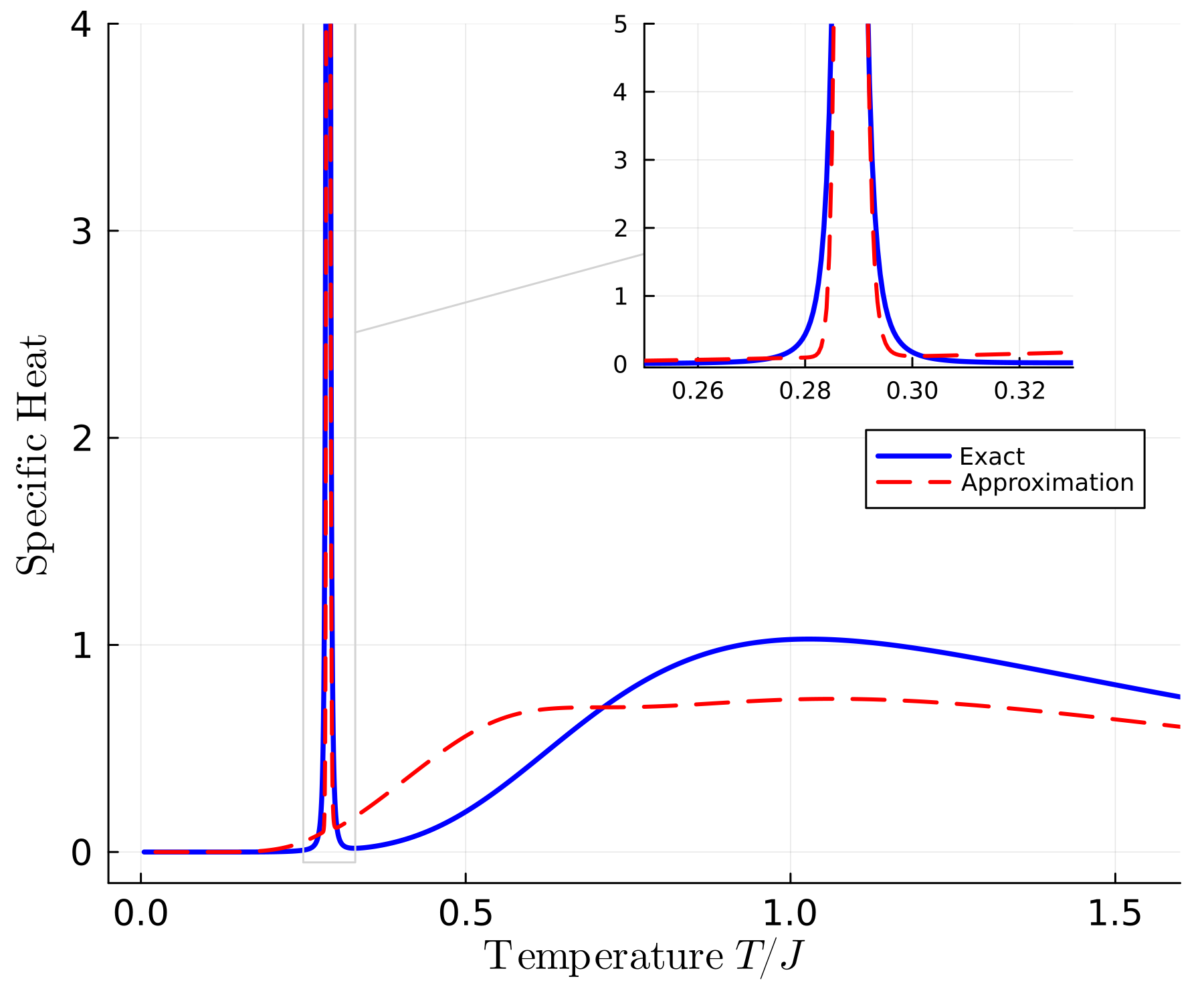}
        \caption{Specific Heat}     
    \end{subfigure}
    \caption{Entropy (a) and specific heat (b) obtained through our approximation (red, dashed line) as compared to the exact solution of the 1D Toblerone model (blue, solid line). We use $J = 2, J_\Delta =1.2, J_R = -1$. We see good agreement between the approximation and the exact solution, with the approximation capturing the physics of the pseudo-transition. The inset of panel (b) shows that the width of the peak in the specific heat is well described in this approximation.}
    \label{fig:ICM_thermo}
\end{figure}

We demonstrate the efficacy of this approximation through a comparison to the one-dimensional ``Toblerone model", for which there is an exact solution to compare to. This comparison of the entropy and the specific heat is shown in \Cref{fig:ICM_thermo}. We see that this approximation not only captures the essential physics (the pseudo-transition), but is quite well behaved. We can be confident with our use of this approximation for the two-dimensional decorated bilayer.

The comparison to the 1D model is non-trivial. There is no particular reason that a correlated cluster should differ significantly between one- and two-dimensions. Whilst there are quantitive differences in \cref{fig:ICM_thermo}, the CCA accurately reproduces the width of the peak in the specific heat, as show in the inset of \Cref{fig:ICM_thermo} panel (b). Further, the deviation, and broadening, of the peak as we move away from the line $J_\perp=0$, is captured in this approximation. More details of this can be found in Ref.~\cite{chapman2024bifurcation}. Indeed, whilst the CCA does not provide quantitatively exact results, we are able to obtain physical accuracy that goes beyond a schematic description. Without a clear analytical path for analysis, the approximation is sufficient for the present purposes.

\vspace{2em}

\section*{References}
\bibliography{refs}

\end{document}